\documentclass[11pt]{article}
\usepackage{moriond,epsfig}
\def\MyPhoto{\psfig{figure=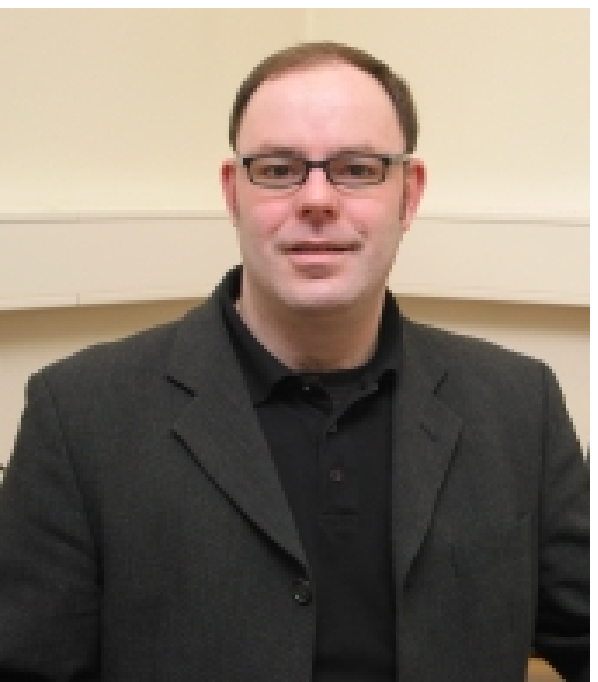,width=35mm}}

\begin{document}
\vspace*{4cm}
\title{Higgs Boson Searches with ATLAS based on 2010 Data}
\author{ M. Schumacher on behalf of the ATLAS collaboration}
\address{Fakult\"at f\"ur Mathematik und Physik,  
Albert-Ludwigs Universit\"at Freiburg, Hermann-Herder-Strasse 3,\\ 
D-79104 Freiburg im Breisgau, Germany}
\maketitle\abstracts{
The results of Higgs Boson searches with the ATLAS detector based on 2010 proton proton 
collision data corresponding to integrated luminosities of up to 39 pb$^{-1}$ are presented. 
Searches for $H \rightarrow \gamma\gamma$, $H\rightarrow W⁺W⁻ \rightarrow 
\ell^+ \nu \ell^-\bar{\nu}$ and $H \rightarrow ZZ \rightarrow \ell^+\ell^-\nu\bar{\nu}/\ell^+\ell^-q\bar{q}$  
in the context of the Standard Model (SM), for $H \rightarrow \tau⁺\tau⁻$ in the context of 
the Minimal Supersymmetric Extension of the Standard Model (MSSM) and for a generic scalar 
at low mass in the vicinity of the $\Upsilon$ resonance decaying to a pair of muons are discussed. 
All observations are in agreement with the expectations from the background-only hypothesis.
Hence exclusion limits at 95\% confidence level are derived.}

\section{Introduction}
Unraveling the mechanism responsible for electroweak symmetry breaking and the
generation of elementary particle masses is one of the great scientific quests
of high energy physics today. The Standard Model (SM) and its supersymmetric 
extensions address this question by the Higgs-Englert-Brout-Gurnalik-Hagen-Kibble mechanism.  The first manifestation of this mechanism is represented 
by the existence of at least one Higgs boson. This motivates the large 
experimental effort for the Higgs boson search in the past, presence and future. 
During the year 2010 
the Large Hadron Collider (LHC) delivered proton proton collisions at a center 
of mass energy of 7 TeV corresponding to an integrated luminosity of 48 pb$^{-1}$ 
to the ATLAS experiment~\cite{Aad:2008zzm}.  These data have been used to search for 
Higgs bosons in the SM and its supersymmetric extensions in a variety of final states. 
For the signal rates, their central values and the 
estimation of the associated systematic uncertainties, which arise from variations in the 
renormalization and factorization scales, the choice of the value of the strong coupling constant 
and the choice of the parton distribution functions of the proton, the recommendations from the 
LHC Higgs cross-section working group have been used~\cite{xs}.
As no hints for the production of Higgs bosons are observed in data, exclusion limits 
at the 95\% confidence  level are derived. In order to do so the profile likelihood 
method~\cite{Cowan:2010js} is used as the test statistic, which allows systematic 
errors to be incorporated in the signal and background predictions as nuisance 
parameters.
As the main result power constrained exclusion limits~\cite{Cowan:2011an} to the signal+background 
hypothesis are derived ($PCL_{S+B}$). The power constraint requires that the confidence level 
for the background-only hypothesis is at least 16\% ($CL_B >$ 0.16). Hence, if the observed 
$CL_{S+B}$ is smaller than the expected median $CL_{S+B}$ minus one standard deviation, the observed 
limit quoted is replaced by $CL _{S+B}^{expected}-1\sigma$. For comparisons with other experiments 
the exclusion limits obtained from $CL_S = CL_{S+B}/CL_{B}$~\cite{cls} are also given 
in~\cite{hgg,hww,hzz,htt,hmm}, in which also details of the individual analyses discussed below 
can be found. 

\section{Searches for the Higgs Boson of the SM}\label{sec:hsm}
\subsection{$H \rightarrow \gamma \gamma~$ with~38~pb$^{-1}$}
\label{subsec:hgg}
\begin{figure}[h!b!t!]
\psfig{figure=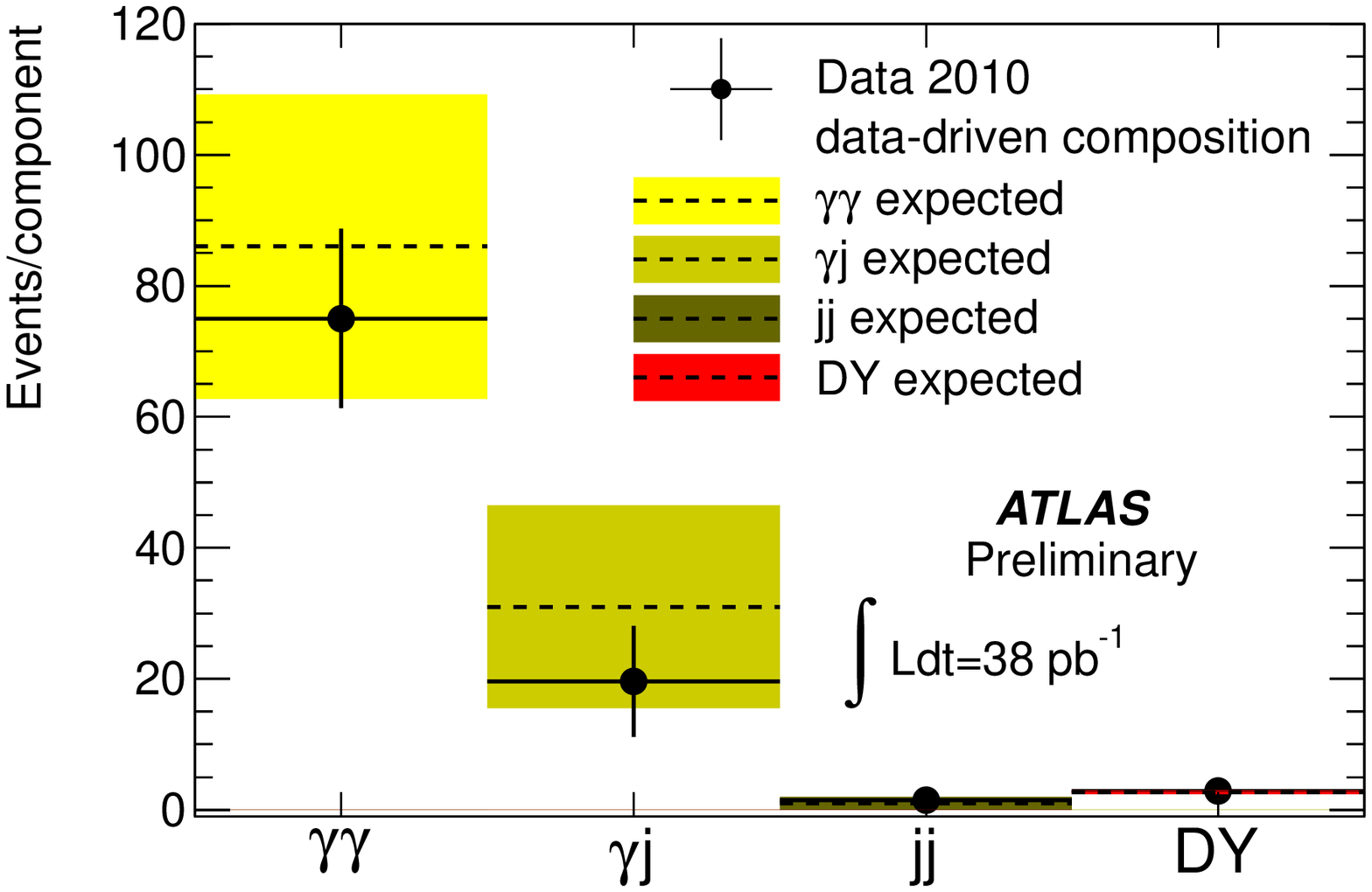,width=5.25cm,height=4.4cm}
\psfig{figure=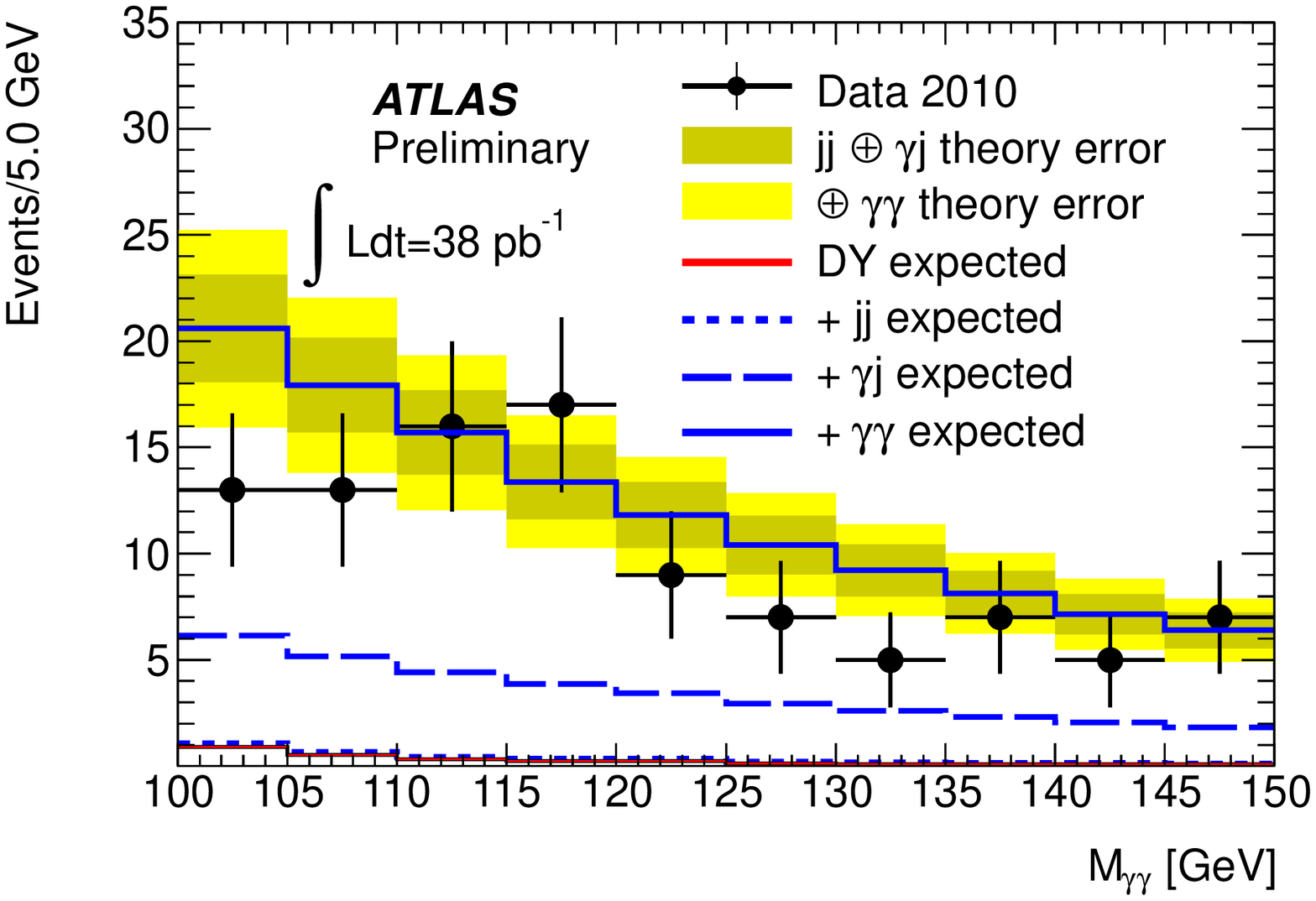,width=5.25cm,height=4.4cm}
\psfig{figure=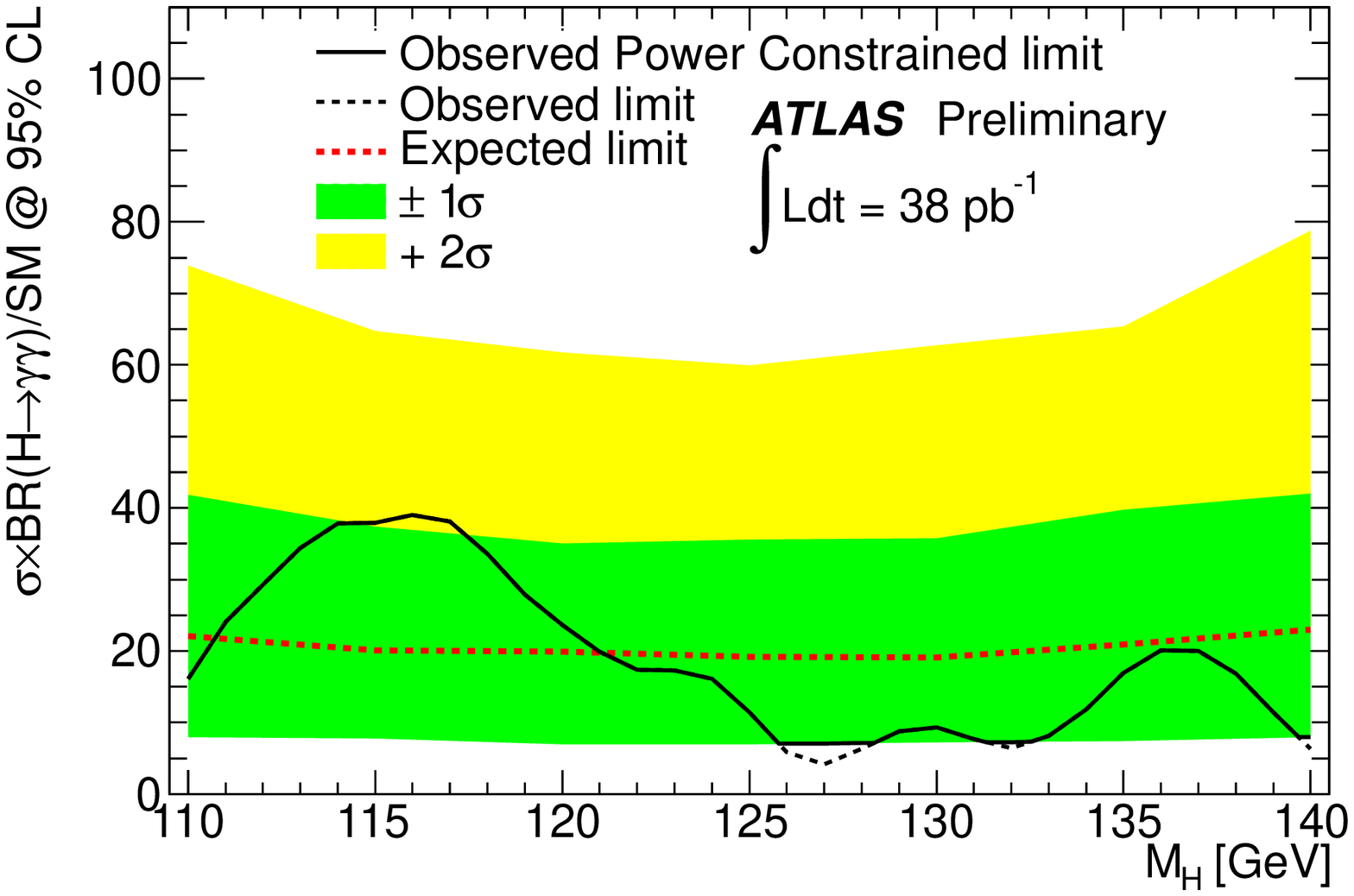,width=5.25cm,height=4.4cm}
\caption{$H\rightarrow \gamma \gamma$: comparison of background prediction from simulation
with the results from the data-driven technique (left), observed invariant di-photon mass 
spectrum with the one predicted from simulation (middle), and excluded signal cross-section with 
respect to the SM prediction (right).\label{fig1}}
\end{figure}

In the mass range from 100 GeV to 140 GeV the decay of the SM Higgs boson into two photons provides 
a very good sensitivity to observe Higgs boson production. The signal topology is characterized by 
two isolated photons with large transverse momentum ($p_\mathrm{T}^{1(2)}>$ 40(25) GeV). 
The reducible 
background arises from photon plus jet(s) and multijet production. These backgrounds are suppressed 
by the excellent capabilities of the ATLAS detector to discriminate photons from jets. The 
irreducible background stems from di-photon production which can be separated from the signal by 
excellent reconstruction of the invariant mass of the di-photon system. The contributions from 
the different background classes have been estimated from data using an iterative double 
sideband method by comparing event yields of loosely and tightly identified photon candidates 
which are isolated or non-isolated. The events yields extracted via this method are in 
good agreement with the prediction from simulations (see Fig.~\ref{fig1} (left)). The invariant 
di-photon mass spectrum after all selection cuts applied is shown in Fig.~\ref{fig1} (middle). 
No significant resonance structure is observed. The background is parametrized via an exponential 
shape with two nuisance parameters (normalization and slope) and no use is made of the MC prediction.
The signal shape is described by the sum of a Crystal Ball function~\cite{cb} plus a Gaussian with a full width at half maximum of 4.4 GeV. The uncertainties on the signal yield are dominated by the 
uncertainty on the inclusive signal cross-section (20\%) and the one on the photon identification
and isolation (10\% each). The width of the hypothetical signal is known to a level of 13\%
from the energy scale and resolution uncertainties for photons.
The expected limits are at the level of 20 times the SM predicted rate (Fig.~\ref{fig1} (right)), 
and the observed exclusion limits lie in the range between 8 and 38 times the SM cross-section 
in the mass range between 110 and 140 GeV.

\subsection{$H \rightarrow W⁺W⁻ \rightarrow \ell^+ \nu \ell^-\bar{\nu}$~ with~35~pb$^{-1}$}
\label{subsec:hww}
\begin{figure}[h!b!t!]
\psfig{figure=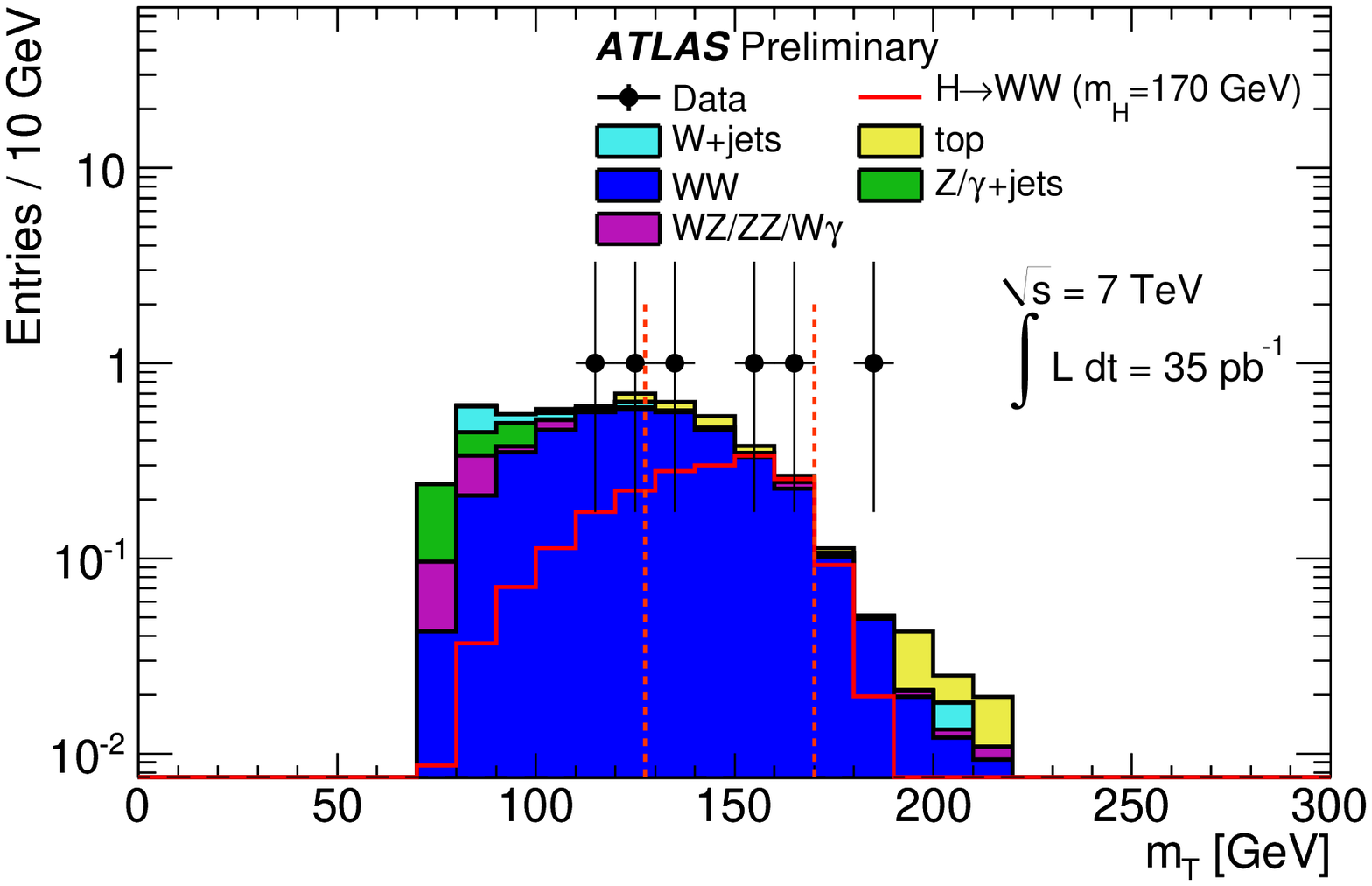,width=5.25cm,height=4.4cm}
\psfig{figure=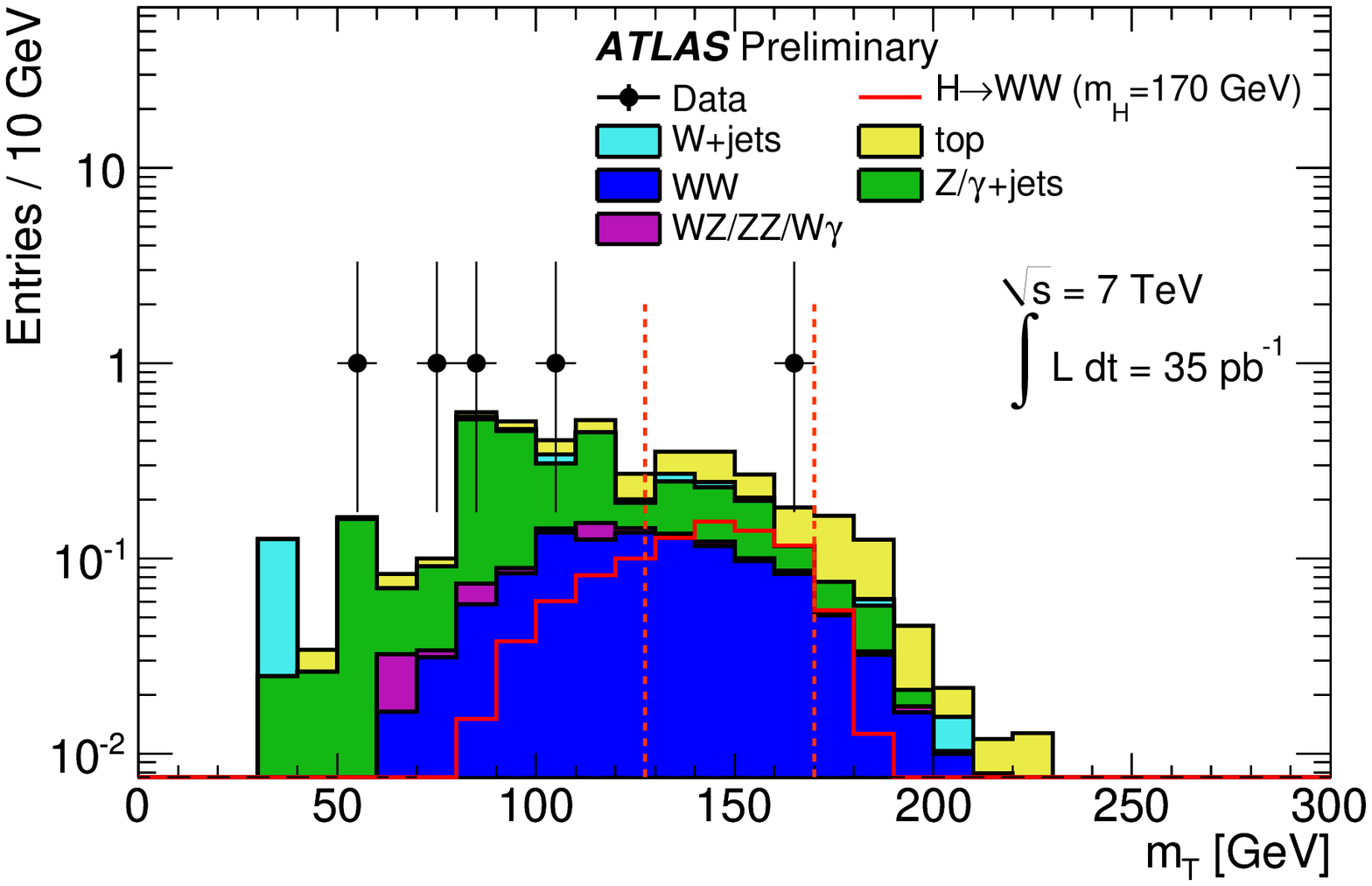,width=5.25cm,height=4.4cm}
\psfig{figure=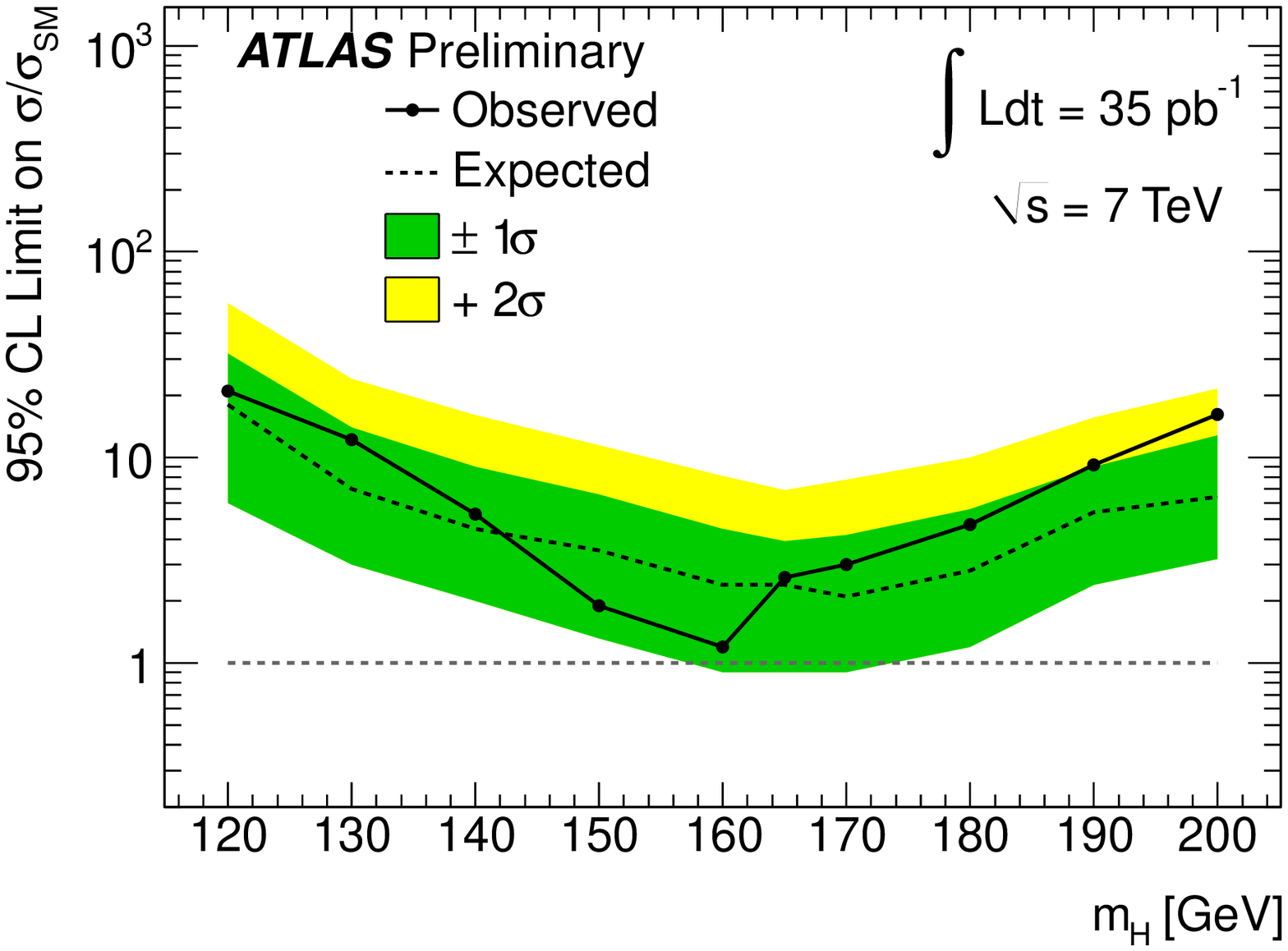,width=5.25cm,height=4.4cm}
\caption{ $H\rightarrow W^+W^-$: observed transverse mass spectrum compared to the background prediction
for zero and one jet analyses (left and  middle), and excluded signal cross-section 
with  respect to the SM prediction (right). \label{fig2}}
\end{figure}

The decay of Higgs bosons into a pair of $W$ bosons yields the highest sensitivity
for an early discovery of a Higgs boson at the LHC especially in the mass range 
around 170 GeV. The decay of each $W$ boson to an electron or muon and the corresponding 
neutrino, which is produced in gluon fusion or weak vector boson fusion, is considered.
The preselection exploits the basic signature of two leptons, which due to spin correlations 
are close in phase space, and significant transverse missing energy (MET) arising from the 
two undetected neutrinos. In order to maximize the sensitivity the analysis is split
into a zero, one and two jet selection, where different additional topological cuts are 
applied in each branch. The uncertainty on the fraction of signal events in each jet topology
is determined by varying renormalization, factorization scales, parton density functions, 
and strong coupling constant in a NNLO+NNLL calculation for Higgs production in gluon fusion. 
Finally a cut on the transverse mass $M_\mathrm{T}$ derived from 
from the lepton momenta and the MET is applied, which 
depends on the Higgs boson mass hypothesis  
($0.75 \cdot M_H < M_{\mathrm{T}} < M_H$). 
The distribution of the transverse mass after all cuts for the zero and one jet analyses 
is shown in Fig.~\ref{fig2} (left and middle). 
The individual background contributions are derived from signal-free control regions in data, which 
are defined by inverting and omitting selection criteria or applying additional requirements to enhance a 
specific contribution.  The extrapolation factors from control regions to signal regions as well 
from one control region to another control region are derived from MC simulated event samples. 
The extrapolation factors are subject to a variety of systematic uncertanties: experimental ones 
from lepton and MET energy scale and resolution uncertainties and theoretical ones from
variation of QCD scales etc.\ (details can be found in~\cite{hww}). As no deviatons from the background 
only hypothesis are observed, exclusion limits on the Higgs boson production 
cross-section are derived (see Fig.~\ref{fig2} (right)). Higgs boson production for a mass of 160 GeV 
with a rate larger than 1.2 times the SM rate can already be excluded. 

\subsection{$H \rightarrow ZZ \rightarrow
\ell^+\ell^-\nu\bar{\nu}/\ell^+\ell^-q\bar{q}$~with~35~pb$^{-1}$}
\label{subsec:hzz}
\begin{figure}[h!b!t!]
\psfig{figure=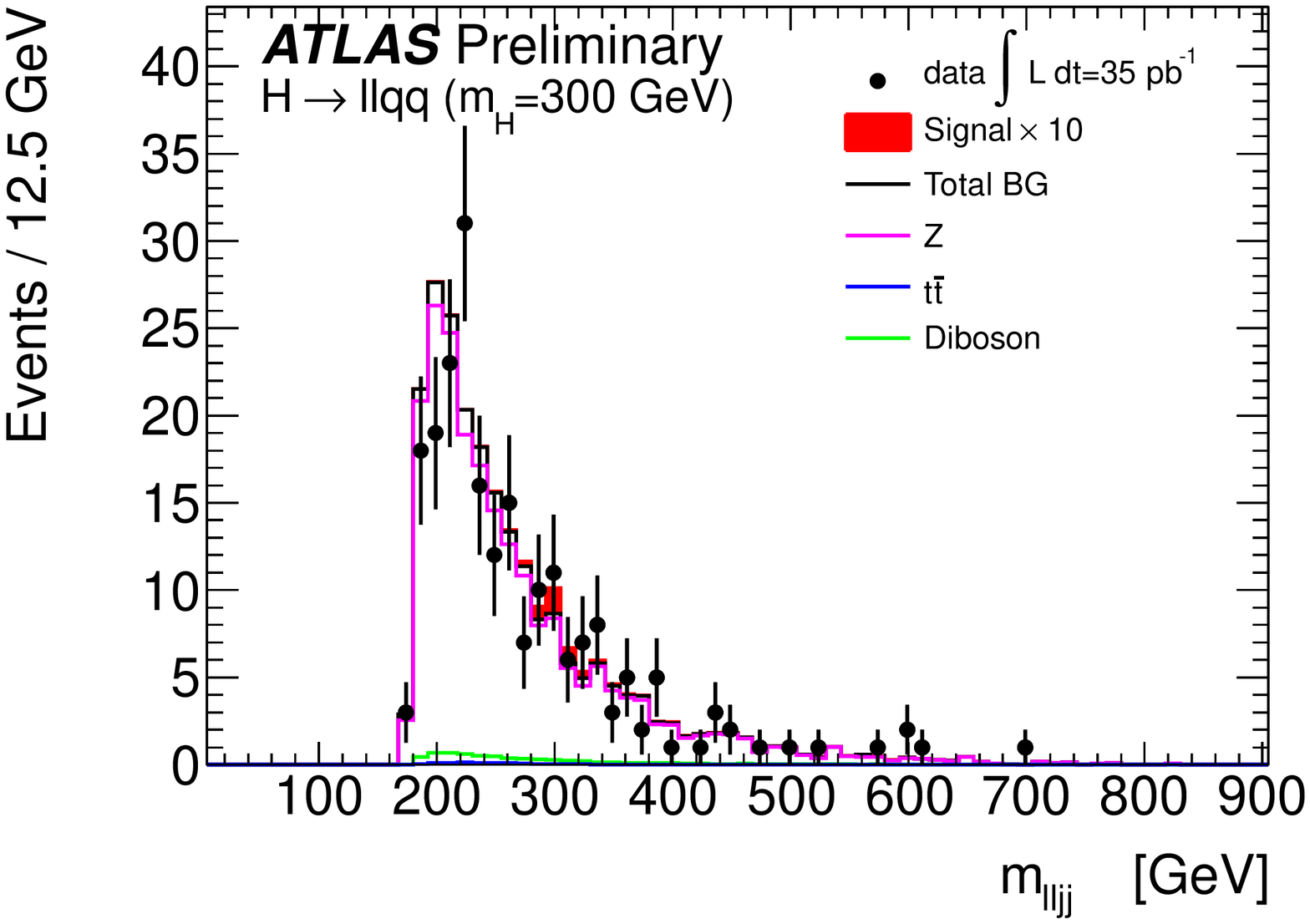,width=5.25cm,height=4.4cm}
\psfig{figure=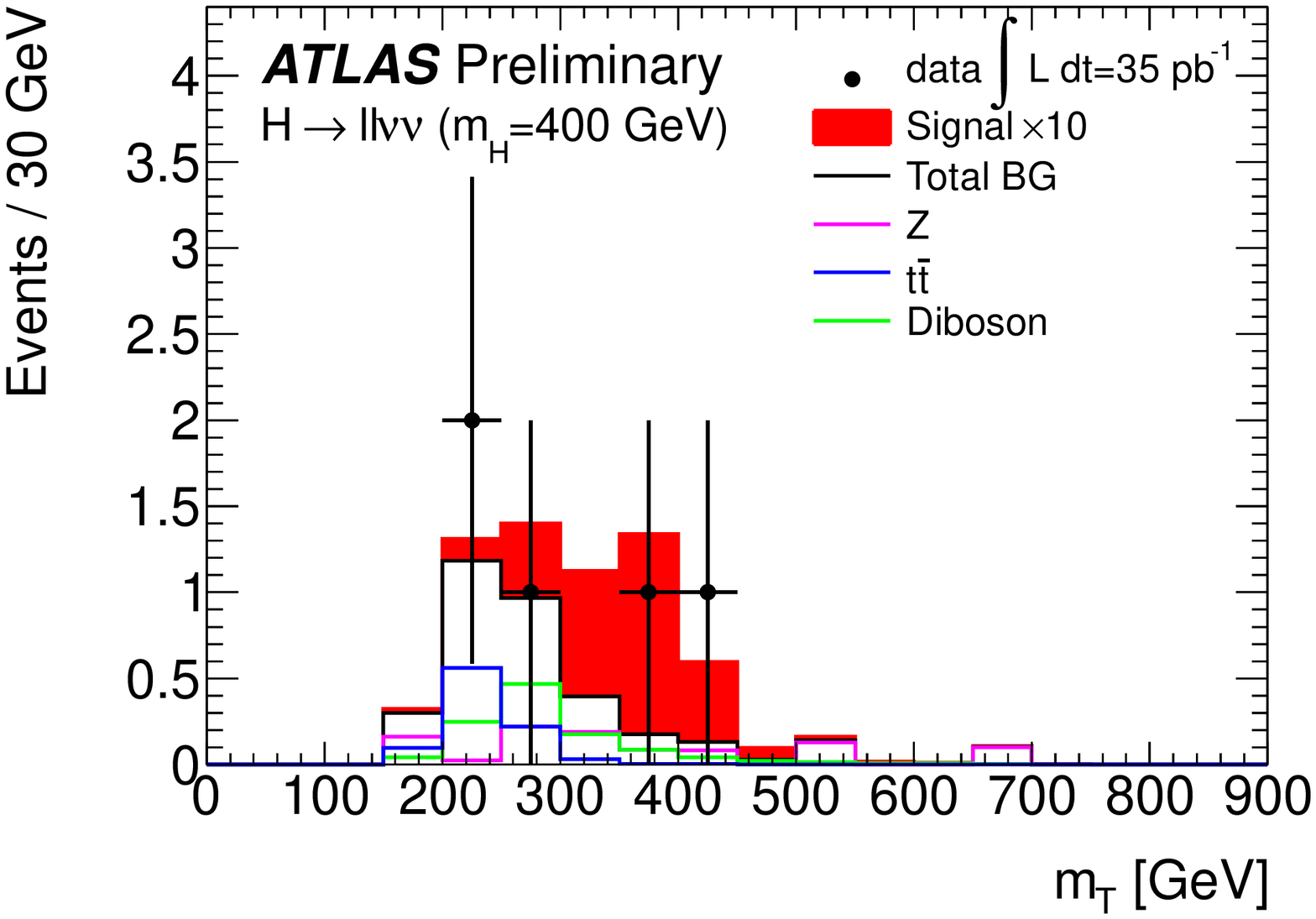,width=5.25cm,height=4.4cm}
\psfig{figure=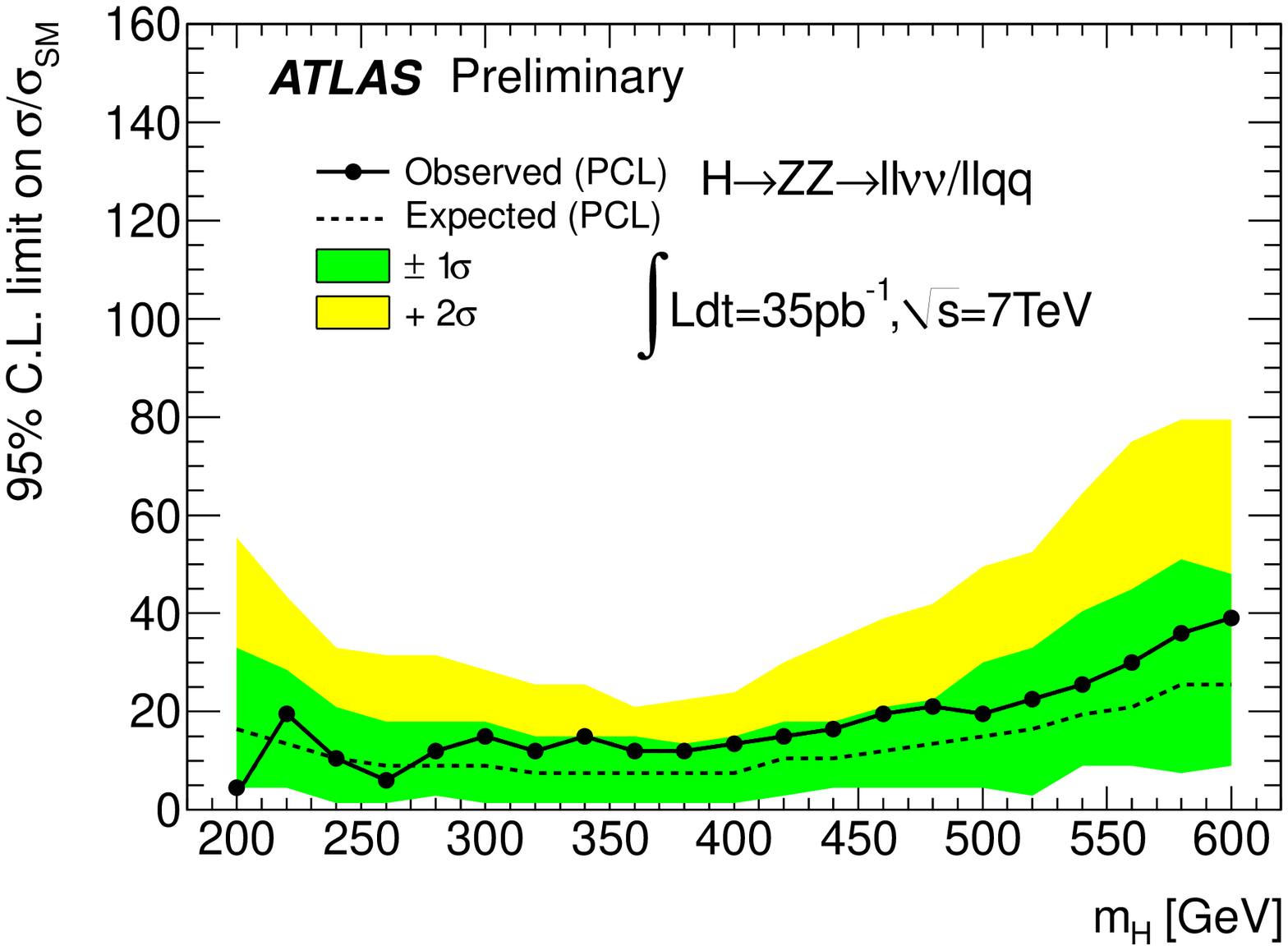,width=5.25cm,height=4.4cm}
\caption{ $H\rightarrow ZZ$: observed  mass spectrum compared to the background 
prediction for the $\ell\ell qq$ final state (left),
observed  transverse mass spectrum compared to the background 
prediction for the $\ell\ell \nu\nu qq$ final state (middle),
and excluded signal cross-section with  respect to the SM prediction (right). \label{fig3}}
\end{figure}

Higgs boson decays into $ZZ \rightarrow \ell^- \ell^+ q\bar{q}$ and
$ZZ \rightarrow \ell^+\ell^-\nu\bar{\nu}$ in the mass range between 
200 and 600 GeV have been searched for. The signal is characterised
by one pair of same flavour, oppositely charged leptons with invariant 
mass consistent with the $Z$ boson and either a pair of jets whose
invariant mass is also consistent with the $Z$ boson mass or large MET 
due to the two neutrinos in the final state. Additional kinematical
cuts are applied to suppress the backgrounds.  The shape and normalisation of the 
expected backgrounds have been confirmed by comparing MC simulated event samples 
with the observed event yield in data  e.g. in sideband regions of 
$M_{\ell^+\ell^-}$ and $M_{q\bar{q}}$ and other control regions.
No deviation from the SM expectation without a Higgs boson are observed 
in the final mass distributions (shown in Fig.~\ref{fig3} (left) and (middle)) and hence 
exclusion limits with repect to the SM production rate are set. Those are in 
the range of 3.5 to 39 times the SM prediction and are currently the most 
stringent exclusion limits for Higgs boson mass hypotheses beyond 300 GeV.

\section{Search for $H \rightarrow \tau\tau$ 
in the MSSM with 36 pb$^{-1}$}\label{sec:htt}
\begin{figure}[h!b!t!]
\psfig{figure=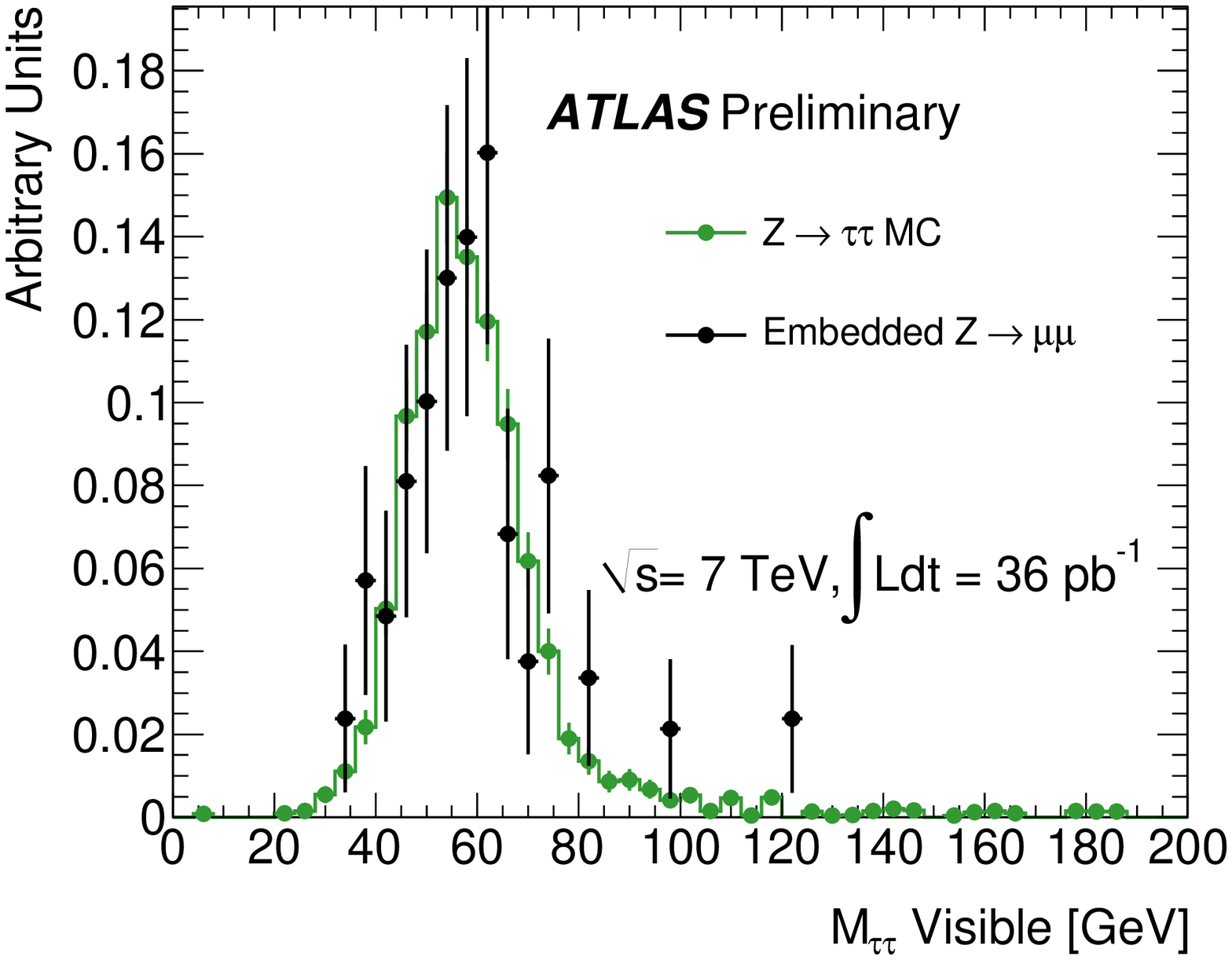,width=5.25cm,height=4.4cm}
\psfig{figure=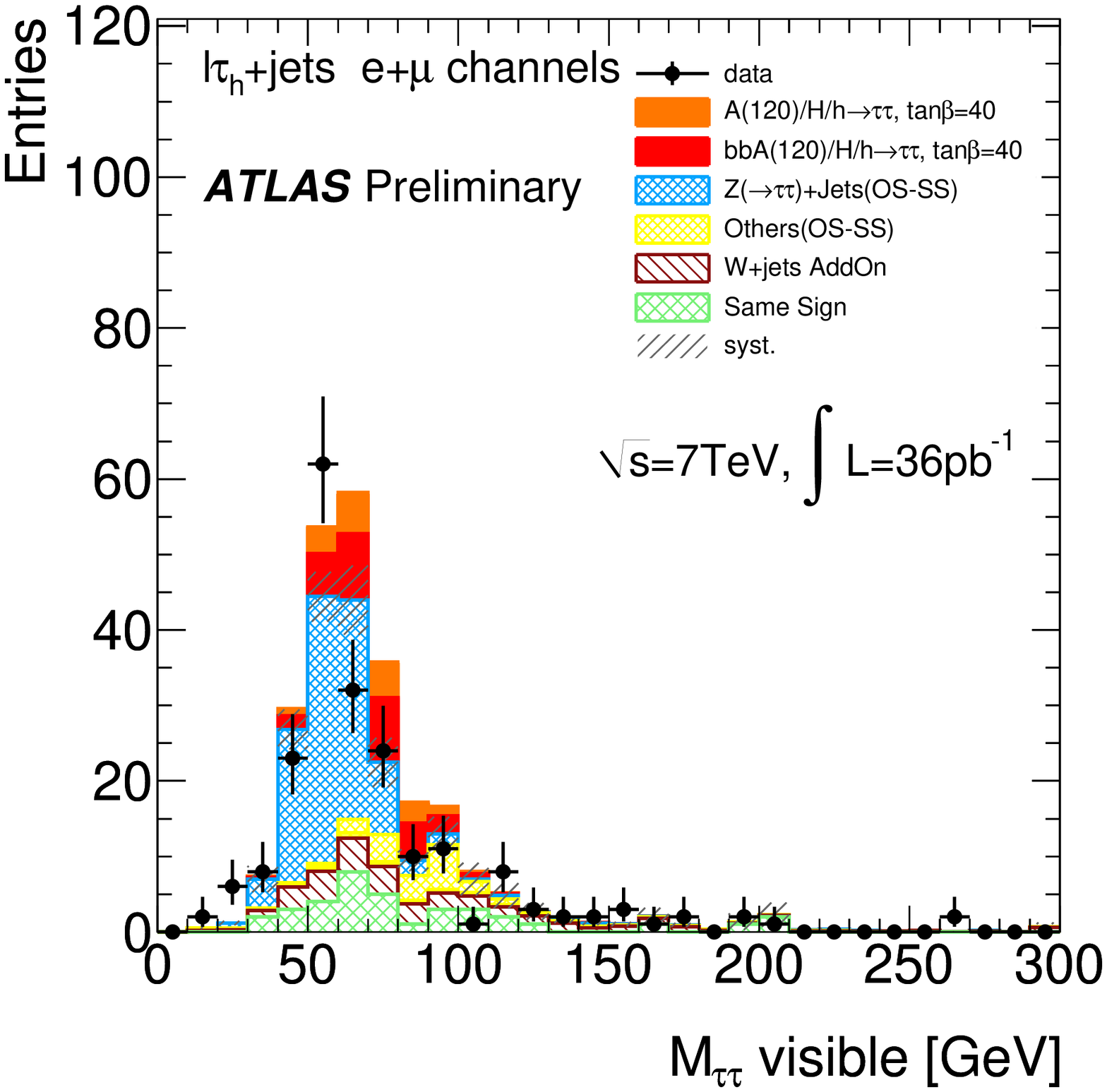,width=5.25cm,height=4.4cm}
\psfig{figure=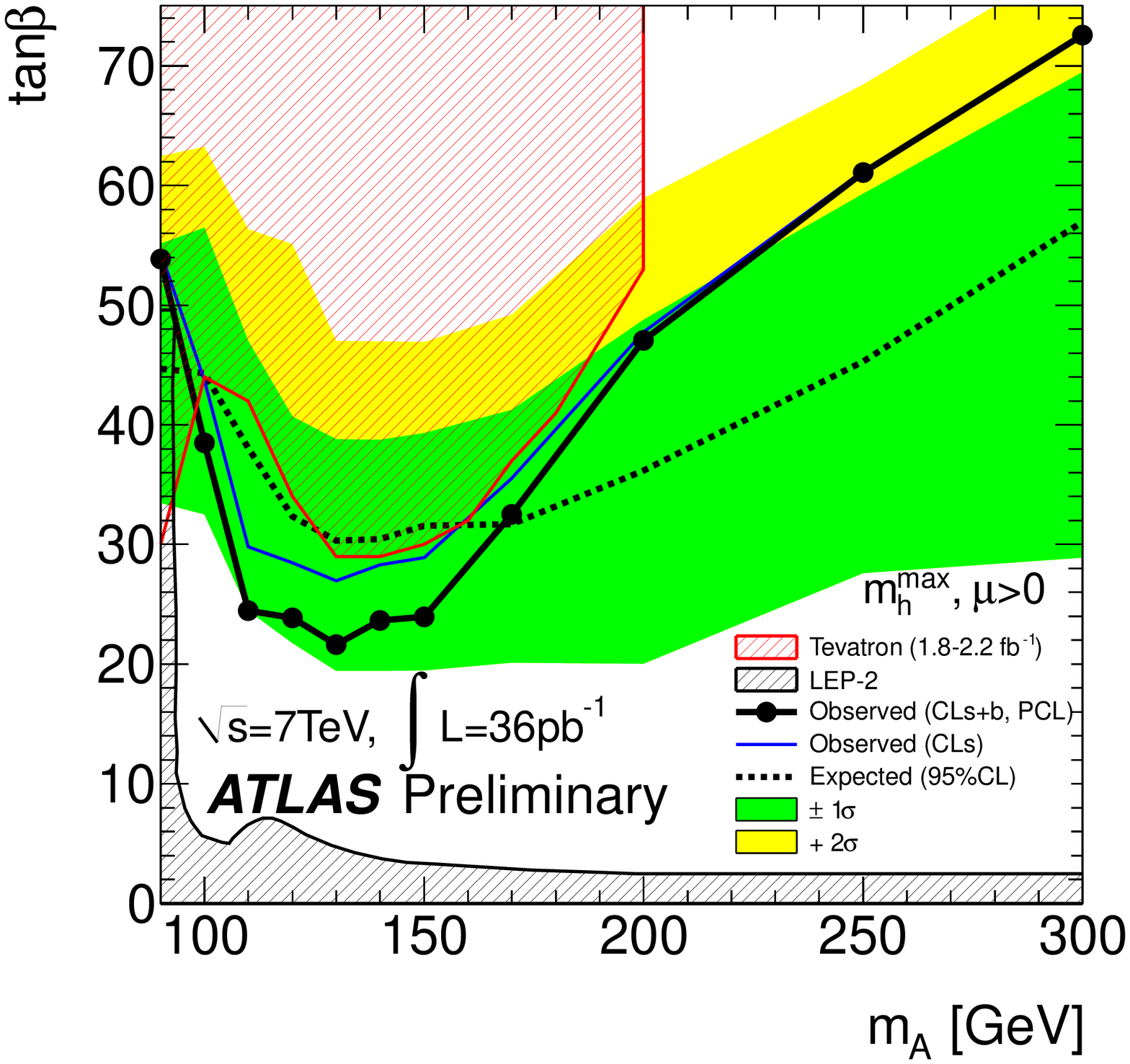,width=5.25cm,height=4.4cm}
\caption{$H\rightarrow \tau\tau$: comparison of the visible mass shape
of embedded $Z\rightarrow\mu\mu$ collision events with simulated $Z\rightarrow \tau\tau$ events (left),
comparison of the visible mass distribution of event selected in data with the prediction 
for all background processes and a hypothetical signal (middle), and excluded parameter space
in the  MHMAX benchmark scenario of the MSSM (right). \label{fig4}}
\end{figure}

The most promising channel for the observation of Higgs bosons in the context of the 
MSSM is the decay into a pair of tau leptons. Production in gluon fusion and
in association with b-quarks have been considered. One tau lepton is assumed
to decay hadronically, the other one leptonically.  The inclusive selection exploits
the signal characteristics by requiring one electron and muon, an oppositely charged hadronically 
decaying tau candidate and significant MET and applies an upper cut on the transverse mass
of the lepton and MET system in order to suppress background from $W$ boson production.
The final discriminant is the invariant mass of the visible tau decay products.  The prediction of the 
mass shape for the irreducible $Z\rightarrow \tau\tau$ background in MC simulated events has been 
confirmed in data, by selecting  $Z\rightarrow \mu\mu$ collision events and replacing the muons 
by tau lepton decays from simulation with the same kinematic properties (see Fig.~\ref{fig4} (left)). 
The backgrounds with fake tau candidates dominated by $W+jet$ production has been estimated mostly 
from data by using the observed mass shape of signal free events with same charge sign of 
electron or muon and tau candidate and determining the ratio of same sign to opposite sign 
events for $W+jet$ 
and multijet production in control regions in data. The normalization of $Z\rightarrow\tau \tau$ 
background and the normalization and shape of other small backgrounds are obtained from MC simulated 
event samples. The dominant systematic uncertainties on shape and normalization of the $\tau \tau$ 
event yield from $Z$ and $H$ boson production arise from the uncertainty in the tau lepton energy scale and 
the jet energy scale its influence on the MET scale and to a lesser extent in the tau lepton 
identification efficiency. The 
final visible mass distribution (see Fig.~\ref{fig4} (middle)) with data-driven background predictions 
compared to the data shows no hint for Higgs boson production.  Hence parameter regions in 
the $M_A$-vs.-$\tan\beta$ plane of the MHMAX benchmark scenario ($\mu > 0$) 
of the MSSM~\cite{Carena:2002qg} can be excluded (see  Fig.~\ref{fig4} (right)). 
At $M_A$ = 130 GeV $\tan\beta$ values above 22 can be excluded. These findings 
extend those published previously by the LEP and Tevatron experiments~\cite{lepmssm,tevmssm}.
\section{Searches for $\phi \rightarrow \mu^+ \mu^-$ 
at low mass with 39 pb$^{-1}$}\label{sec:hmm}
\begin{figure}[h!b!t!]
\psfig{figure=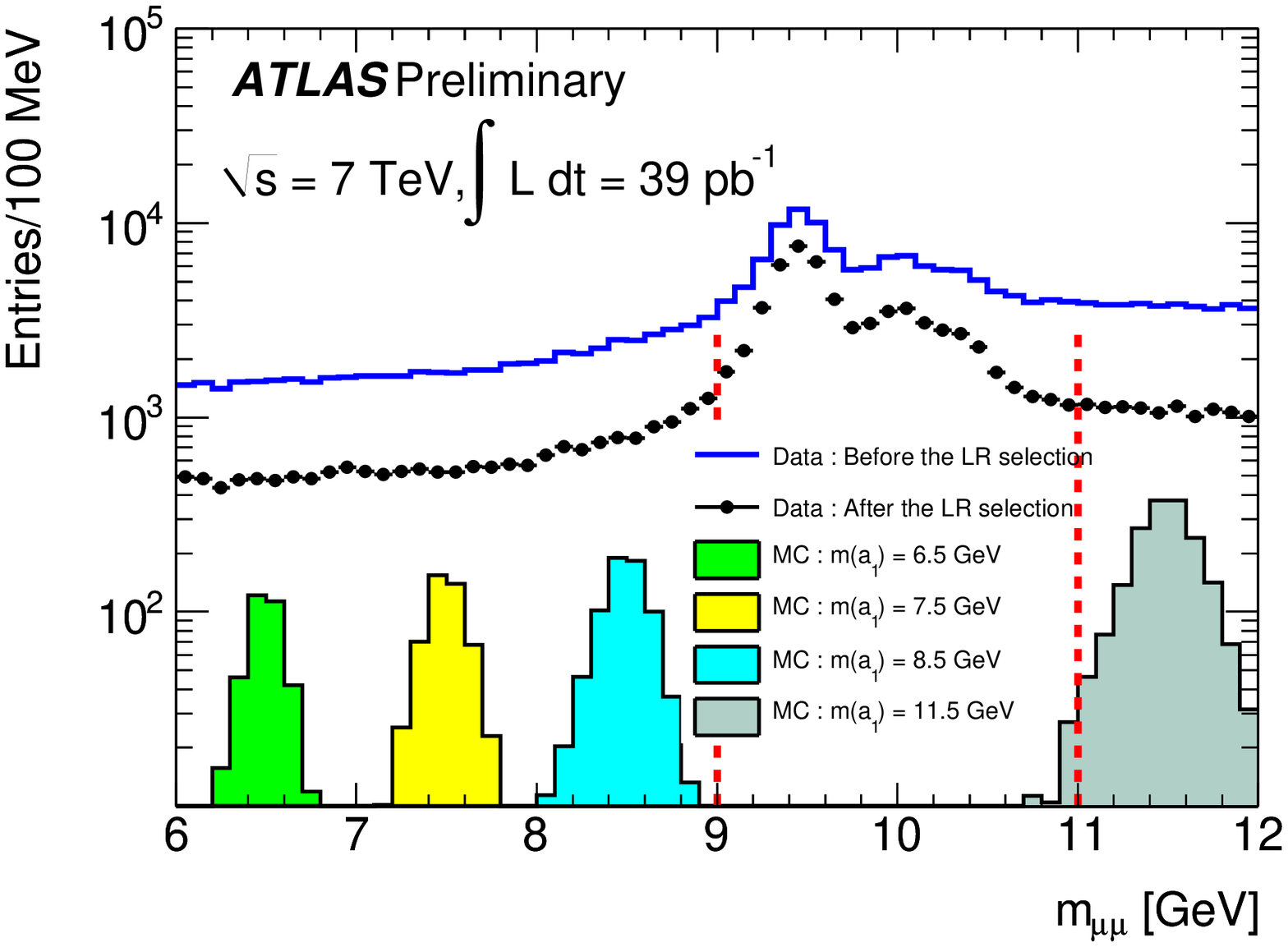,width=5.25cm,height=4.4cm}
\psfig{figure=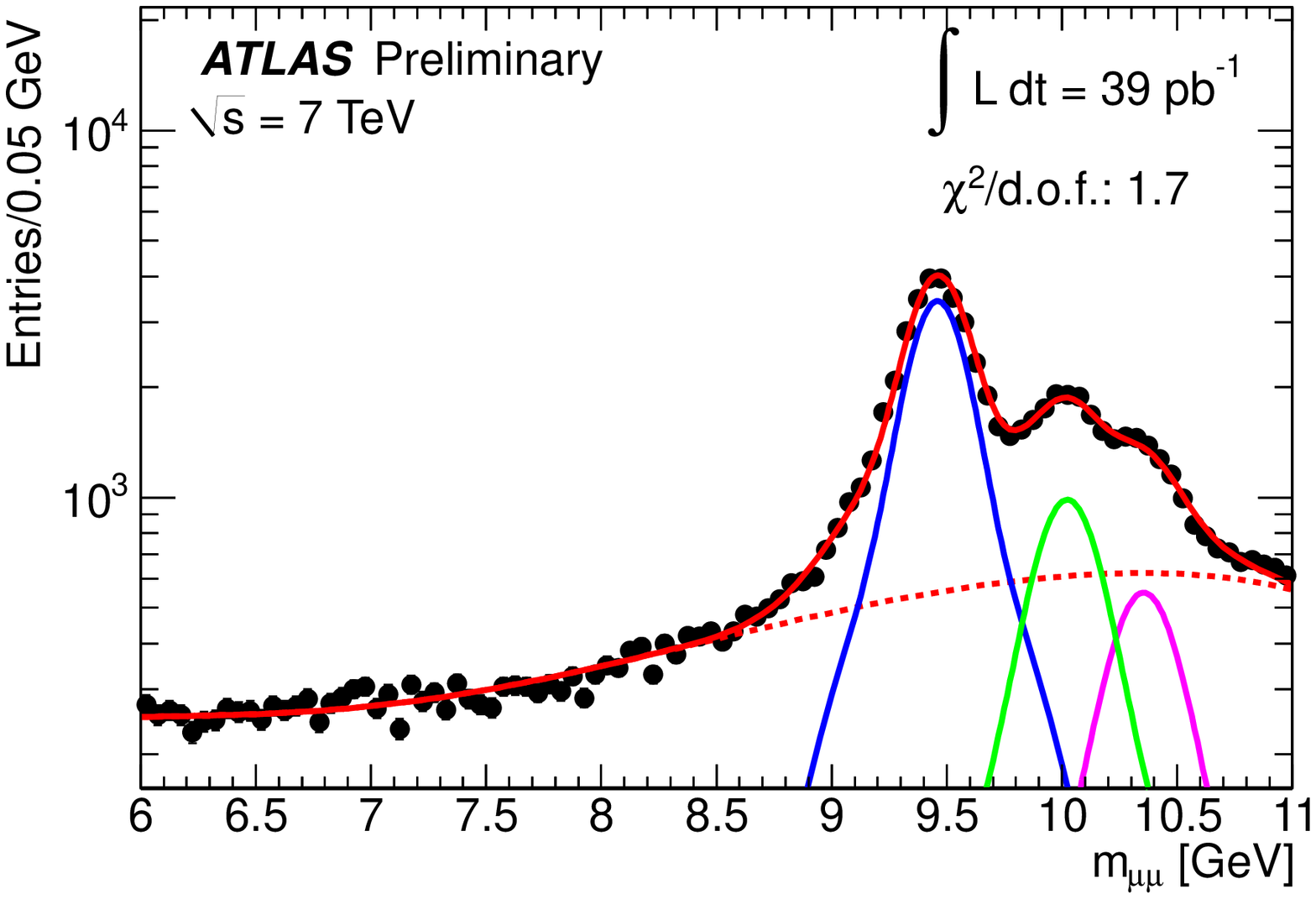,width=5.25cm,height=4.4cm}
\psfig{figure=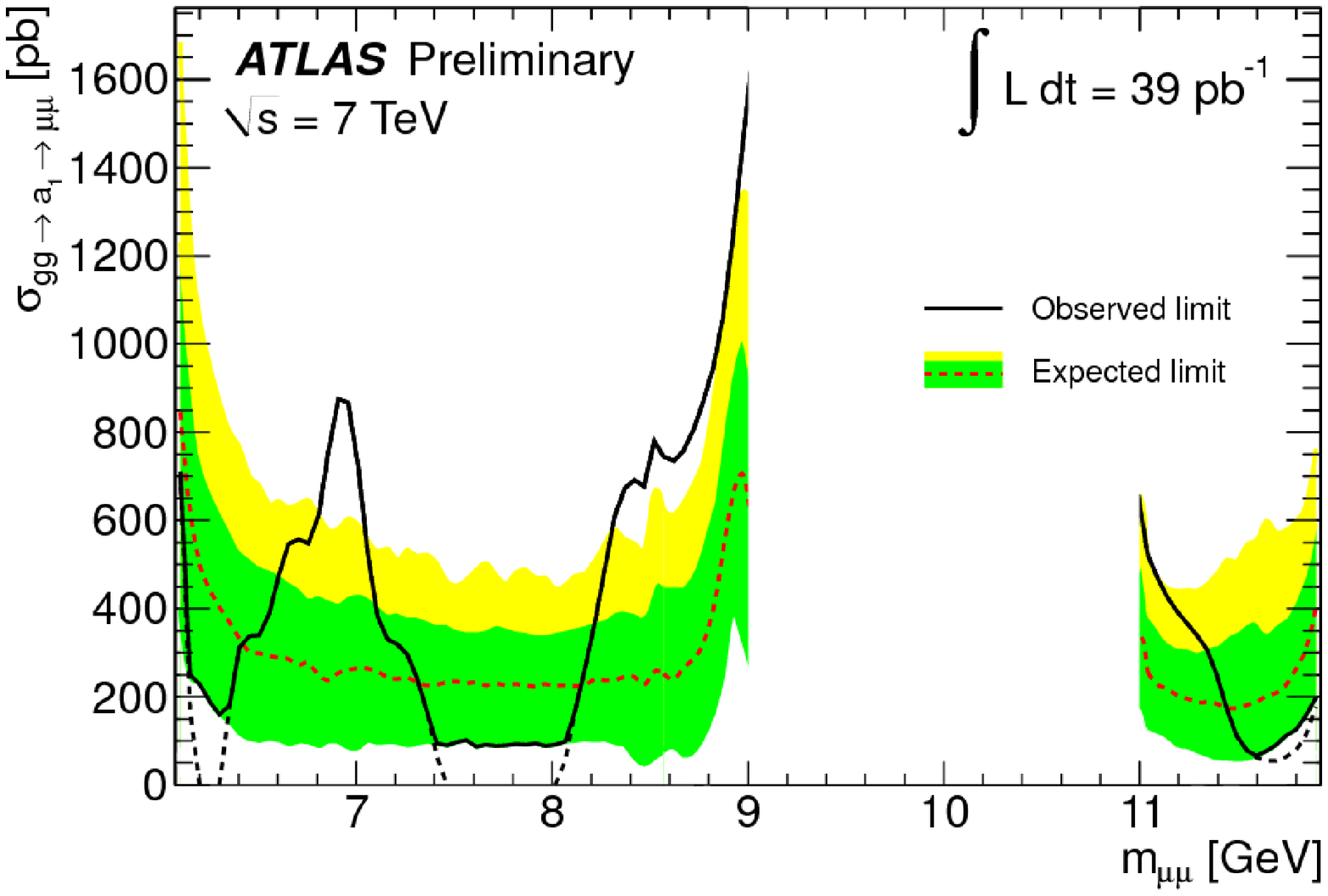,width=5.25cm,height=4.4cm}
\caption{$\phi \rightarrow \mu\mu$: observed invariant di-muon mass 
spectrum (left), result of the background-only hypothesis fit to the invariant di-muon mass spectrum (middle), and cross-section times braching ratio limit(right).\label{fig5}}
\end{figure}

In extensions of the MSSM either via additional singlets as in the NMSSM 
(see e.g.~\cite{Maniatis:2009re}) or via allowing for complex parameters 
in the MSSM, yielding additional sources of CP-violation, the existence of 
a low mass Higgs boson in the vicinity and below the masses of the $\Upsilon$ resonances 
is not completely excluded. A search for a generic scalar $\phi$ in the mass range 
from 6 to 9 and 11.0 to 12.5 decaying to a pair of muons produced in gluon fusion has been 
performed.  After selecting events with two muons with a transverse momentum exceeding 
4 GeV a likelihood ratio selection is applied. The probability density functions of the input 
variables for the signal and background hypothesis are derived from data itself by selecting 
events outside the search region: i.e. 9 to 11 GeV for the signal hypothesis, which has been 
confirmed to be kinematically identical to $\Upsilon$ production, and for the background hypothesis 
from events with $M_{\mu\mu}$ below 6 and above 11.5 GeV. The final mass distribution after 
applying a cut on the likehood ratio is shown in Fig.~\ref{fig5} (left). 
The uncertainty on the expected signal yield is estimated to be 70\% for a signal mass
of 6 GeV and 28\% for a signal mass of 11 GeV, which is dominated by the uncertainty
on the kinematical acceptance. The continuum background 
is parametrised by a forth order polynomial, where all parameters are nuisance parameters.
The signal and the $\Upsilon$ resonances are modelled by a double Gaussian probability density function,
where the masses are fixed to the hypothetical signal mass and the world averages for $\Upsilon$
masses, respectively. The width and fraction of the two Gaussians for the $\Upsilon$(1S) 
resonance are nuisance parameters. For the other $\Upsilon$ resonances and the signal resonance
the widths and fractions are obtained using a linear dependence of the mass resolution on 
the resonance mass. All normalisations are left floating in the fits. The cross-section times 
branching ratio limit is shown in Fig~\ref{fig5} (right). Production rates down to 200 pb 
can be excluded.
\section{Conclusions}
LHC and ATLAS performed very well during their operation in 2010. However searches 
in a data sample corresponding to an integrated luminosity of up to 39\, pb$^{-1}$ have 
not revealed any hint for Higgs boson production at LHC yet. So far mostly simple cut based 
selections have been performed and the major backgrounds in most analyses have been estimated or 
at least being confirmed using data-driven methods. In the search for $H\rightarrow \gamma\gamma$
a sensitivity of 20 times the SM Higgs boson production rate has been achieved. 
The $H\rightarrow WW$ search allows to exclude already a production rate of 1.2 times the SM rate 
at a mass of 160 GeV.
The search for $H\rightarrow ZZ$ decays yield the world's best limits for a Higgs boson mass 
above 300 GeV up to now. The sensitivity of the search for neutral Higgs bosons of the MSSM via decays 
to a pair of tau leptons already supercedes the findings obtained at the Tevatron. Collecting a data set 
corresponding to 4 fb$^{-1}$ at  LHC at a center of mass energy of 7 TeV will allow to exclude Higgs boson
mass hypotheses in the SM down to the LEP limit of 114.4 GeV~\cite{lepsm} assuming that no deviation 
from the background-only hypothesis is observed~\cite{sensitivity}. Higgs boson hunters in all 
experiments hope that a different scenario is realised in nature.


\section*{References}
\begin{thebibliography}{99}
\bibitem{Aad:2008zzm}
G.~Aad {\it et al.} [ ATLAS Collaboration ],
JINST {\bf 3 } (2008) S08003.
\bibitem{xs} LHC Higgs Cross-Section Working Group (S. Dittmaier {\it et al.}),
{\it Handbook of LHC Higgs Cross-Sections: 1. Inclusive Observables},
CERN-2011-002, arXiv:1101.0593.
\bibitem{Cowan:2010js}
G.~Cowan, K.~Cranmer, E.~Gross, O.~Vitells,
Eur.\ Phys.\ J.\  {\bf C 71 } (2011)  1554.
\bibitem{Cowan:2011an}
G.~Cowan, K.~Cranmer, E.~Gross, O.~Vitells,
{\it Power-Constrained Limits}, arXiv:1105.3166.
\bibitem{cls}
A.~L.~Read, J.\ Phys.\  {\bf G 28 } (2002)  2693-2704;
A.~L.~Read,
CERN-OPEN-2000-205.
\bibitem{hgg} ATLAS Collaboration,
{\it Search for the Higgs boson in the diphoton final state in 38 pb$^{-1}$ of data recorded with the ATLAS detector at $\sqrt{s}$=7 TeV}, ATLAS-CONF-2011-025.
\bibitem{hww} ATLAS Collaboration, {\it Higgs Boson Searches using the 
$H \rightarrow W^+W^- \rightarrow \ell^+\nu \ell^-\bar{\nu}$ Decay Mode with 
the ATLAS Detector at 7 TeV}, ATLAS-CONF-2011-005.
\bibitem{hzz} ATLAS Collaboration, {\it Search for a Standard Model Higgs 
Boson in the Mass Range 200-600 GeV in the Channels 
$H \rightarrow ZZ \rightarrow \ell^+\ell^- \nu \bar{\nu}$ and 
$H \rightarrow ZZ \rightarrow \ell^+ \ell^- q \bar{q}$ with the ATLAS Detector},
ATLAS-CONF-2011-026.
\bibitem{htt} ATLAS Collaboration, {\it Search for neutral MSSM Higgs bosons 
decaying to $\tau^+ \tau^-$ pairs in proton-proton collisions at $\sqrt{s}$
=7 TeV with the ATLAS Experiment}, ATLAS-CONF-2011-024.
\bibitem{hmm} ATLAS Collaboration,
{\it Search for NMSSM light CP-odd Higgs a1 with $\mu^+ \mu^-$ final states 
with pp collisions at $\sqrt{s}$  =7 TeV in ATLAS}, ATLAS-CONF-2011-020.
\bibitem{cb} M. J. Oreglia,  {\it A Study of the Reactions $\Psi^\prime \rightarrow \gamma \gamma \Psi$}, Ph.D. Thesis, SLAC-R-236 (1980). 
\bibitem{Carena:2002qg}
M.~S.~Carena, S.~Heinemeyer, C.~E.~M.~Wagner, G.~Weiglein, Eur.\ Phys.\ J.\  {\bf C 26 } (2003) 601.
\bibitem{lepmssm}
 S.~Schael {\it et al.} [ALEPH, DELPHI, L3 and OPAL Collaborations and LEP Working Group for Higgs Boson Searches],
Eur.\ Phys.\ J.\  {\bf C 47 } (2006)  547-587.
\bibitem{tevmssm}
D.~Benjamin {\it et al.} [Tevatron New Phenomena \& Higgs Working Group, CDF and D0 Collaborations],
{\it Combined CDF and D0 upper limits on MSSM Higgs boson production in tau-tau final states with up to 2.2 fb-1}, arXiv:1003.3363.
\bibitem{Maniatis:2009re}
M.~Maniatis, Int.\ J.\ Mod.\ Phys.\  {\bf A25 } (2010)  3505-3602.
\bibitem{lepsm} 
R.~Barate {\it et al.}[ALEPH and DELPHI and L3 and OPAL Collaborations and  LEP Working Group for Higgs boson searches],
Phys.\ Lett.\  {\bf B565 } (2003)  61-75.
\bibitem{sensitivity} ATLAS Collaboration,
{\it Further investigations of ATLAS Sensitivity to Higgs Boson Production in different assumed 
LHC scenarios}, ATL-PHYS-PUB-2011-001.
\end{thebibliography}
\end{document}